\documentclass[prb,showpacs,twocolumn]{revtex4}
\usepackage{graphicx}
\begin{document}

\vspace{1cm}

\title[Anisotropic dielectric function ...]
{Anisotropic dielectric function in polar nano-regions of relaxor
ferroelectrics
  }

\date{\today}
\author{J. Hlinka, T. Ostapchuk,  D. Noujni,  S. Kamba, and J. Petzelt}
\affiliation{Institute of Physics, Academy of Sciences of the Czech
Republic,
   Na Slovance 2, 18221 Praha 8, Czech Republic}

\begin{abstract}
The paper suggests to treat the infrared reflectivity spectra of
single crystal perovskite relaxors as fine-grained ferroelectric
ceramics: locally frozen polarization makes the dielectric function
strongly anisotropic in the phonon frequency range and the random
orientation of the polarization at nano-scopic scale requires to
take into account the inhomogeneous depolarization field. Employing
a simple effective medium approximation (Bruggeman symmetrical
formula) to dielectric function describing the polar optic modes as
damped harmonic oscillators turns out to be sufficient for
reproducing all principal features of room temperature reflectivity
of PMN. One of the reflectivity bands is identified as a geometrical
resonance entirely related to the nanoscale polarization
inhomogeneity. The approach provides a general guide for systematic
determination of the polar mode frequencies split by the
inhomogeneous polarization at nanometer scale.
 \end{abstract}


\maketitle

In recent years, there has been enormous effort in studying single
crystals  with intrinsic nanoscopic inhomogeneity, since such
materials  often show a very interesting properties. It was even
proposed that the clustered, inhomogeneous states encountered for
example in high-Tc cuprates, CMR manganites, nickelates, cobaltites,
diluted magnetic semiconductors or ferrolectric relaxors, should be
considered as a new paradigm in condensed matter
physics.\cite{Dagotto} In case of relaxors, the peculiar dielectric
properties of relaxor materials were associated with the presence of
small polar clusters - so called polar nano-regions (PNR's)- already
in the pioneer work of Burns and Dacol\cite{Burns}. However, because
of their small size and random nature, we still lack a clear
understanding of their size distribution, thickness and roughness of
their boundaries, their connectivity, shape anisotropy, distribution
of the associated dipolar moments, their fractal self-similarity,
their dynamics, and so on. PNR's are often represented as small
islands submerged in a non-polar matrix, possibly appearing and
disappearing again in time. On the other hand, the recent
piezoelectric scanning microscopy investigations\cite{Aveiro} of the
surface of PbTiO$_3$-doped relaxors rather invoke a picture of a
fine, hierarchical and essentially static "nanodomain" structure. It
strongly suggests that the common perovskite relaxors are actually
quite densely filled by quasi-static polar nano-regions, and that
the former picture with prevailing non-polar matrix can perhaps be
appropriate only at high temperatures around the so called Burns
temperature.\cite{Burns,Td}

 Throughout this paper we will have in mind common perovskite relaxors like
Pb(Mg$_{1/3}$Nb$_{2/3}$)O$_3$ (PMN), Pb(Zn$_{1/3}$Nb$_{2/3}$)O$_3$
(PZN), Pb(Sc$_{1/2}$Ta$_{1/2}$)O$_3$ (PST), (Pb,La)(Zr,Ti)O$_3$
(PLZT) and similar systems. Various experiments show that the
dipolar moments in these
 relaxors are caused mainly by ionic off-center
displacements. It is difficult to get reliable information about the
directional distribution of these displacements, but the {\em
amplitude} of the relevant ion displacements (eg. Pb in PMN) is
quite well defined,\cite{Vakhushev, spherical} and it is of the same
order as in normal ferroelectrics. We will assume that these local
displacements are more or less parallel within each PNR (one would
be in trouble how to define PNR if it were not the case) and that
the PNR's are at the time scale of our interest essentially
static,\cite{static} (i.e. the ions vibrate around their displaced
but fixed positions, except perhaps for those at PNR boundaries).
Under such conditions, the homogeneous frozen polarization (dipole
moment density) $P_{\rm F}$ can be well defined within each PNR, as
well as the locally homogenous dielectric function ${\bf
\epsilon}(\omega)$, describing the contribution of polar vibrations
inside a given PNR.

It is obvious that the cubic (and in harmonic approximation
isotropic) environment of ions in perovskite structure is broken by
their off-center displacements. Since the displacements are aligned
within a given PNR, parallel ($\parallel P_{\rm F}$) and
perpendicular ($\perp P_{\rm F}$) ionic fluctuations are strongly
inequivalent. In fact, it is quite likely that the parallel
fluctuations feel a more stiff potential, as in the case of usual
ferroelectrics. Within a given PNR, one may thus expect that the
ionic contributions make the ${\bf \epsilon}(\omega)$ tensor
strongly anisotropic. For simplicity, we will assume that the
anisotropy is uniaxial, so that the ${\bf \epsilon}(\omega)$ tensor
has only two principal components, $\epsilon_{\parallel}$ and
$\epsilon_{\perp}$ (parallel and perpendicular to $ P_{\rm F}$). The
aim of this paper is to demonstrate that infrared (IR) reflectivity
spectra of common perovskite relaxors can be rather well understood
by taking into account this anisotropy.

In the view of the above considerations, it seems reasonable to
analyze the influence of PNR on the polar phonon modes under the
following simplifications:

\begin{itemize}
\item  PNR's are frozen at phonon frequencies
\item  PNR size  is much smaller than the IR  wavelength
\item  Volume of the sample is fully covered by PNR's
\item  Shapes of  PNR's are roughly spherical
\item  $P_{\rm F}$ and ${\bf \epsilon}(\omega)$ within a given PNR are
homogeneous
\item   ${\bf \epsilon}(\omega)$ has uniaxial anisotropy (principal axis
$\parallel P_{\rm F}$)
\item  Orientations of principal axe are random
\item  All PNR's have the same ${\bf \epsilon}(\omega)$
\end{itemize}

It is known that the reflectivity calculations for the samples with
dielectric inhomogeneities (spatial fluctuations of dielectric
function) at scales smaller than the wavelength of the probing
radiation can be performed within the effective medium
approach.\cite{EMAIR} In this approach, the medium is fully
characterized by a homogeneous (averaged) effective dielectric
function, $\epsilon_{\rm eff}$, which could be in principle
evaluated directly from the original heterogenous dielectric
function. In view of the simplifications already made, it seems
sufficient to use here a rough approximation, known as the effective
medium approximation (Bruggeman symmetrical formula for binary
composites), where $\epsilon_{\rm eff}$ is given by the implicit
equation\cite{EMA}
\begin{equation}
 \frac{2}{3} \frac{\epsilon_{\perp} -\epsilon_{\rm eff}}{\epsilon_{\perp}
+2\epsilon_{\rm eff}} + \frac{1}{3} \frac{\epsilon_{\parallel}
-\epsilon_{\rm eff}}{\epsilon_{\parallel} +2\epsilon_{\rm eff}} = 0
\end{equation}
The IR reflectivity of a thick (opaque) sample is then evaluated
from the standard formula
\begin{equation}
R= \left | \frac{\sqrt{\epsilon_{\rm eff}} -1 }{\sqrt{\epsilon_{\rm
eff}} +1 } \right |^2~.
\end{equation}
As a matter of fact, we exploit here the equivalence\cite{Stroud} of
our model with that of a dense random binary composite of spherical
particles with  isotropic dielectric functions equal to
$\epsilon_{\parallel}$ and $\epsilon_{\perp}$, and relative
appearance of 1:2 volume ratio.\cite{Stroud}

It can be expected that the principal polar modes of PNR's in
perovskite relaxors are those originating from the 3 F$_{1u}$ polar
modes of the parent cubic structure. The uniaxial anisotropy induced
by the frozen polarization split each of these triply degenerate
modes in an A$_1$+E$_1$ pair (we assume a strong anisotropy
 limit leading to C$_{\infty v}$ symmetry for any $P_{\rm F}$ direction).
 The principal components of the dielectric tensor can be thus
conveniently parameterized assuming the factorized
form\cite{Prosandeev} for generalized damped harmonic oscillator
model:
\begin{equation}
\epsilon_{\parallel}= \epsilon_{\parallel, \infty} \prod_{j=1}^3
\frac{ ( \omega_{{\rm A_{1},LO}j}^2 -\omega^2 -i \omega \gamma_{{\rm
A_{1},LO}j} ) }{ (\omega_{{\rm A_{1},TO}j}^2 -\omega^2 -i \omega
\gamma_{{\rm A_{1},TO}j} )},
\end{equation}
where $\omega_{{\rm A_{1},TO}j}$ and $\omega_{{\rm A_{1},LO}j}$ is
respectively the transverse and the longitudinal frequency of the
$j$-th mode polarized along the local polarization direction,
$\gamma_{{\rm A_{1},TO}j} $ and $\gamma_{{\rm A_{1},LO}j} $ are the
corresponding damping parameters and $\epsilon_{\parallel, \infty}$
is the corresponding component of the high frequency permittivity
tensor. The $\epsilon_{\perp}$ tensor component determined by E$_1$
modes is defined analogously.

As an example, let us now apply the model (eqs. (1)-(3)) to our room
temperature reflectivity data of PMN,\cite{Prosandeev} reproduced in
Fig.~1. The real and imaginary part of the resulting effective
permittivity is shown in  Fig.~2 and it is obviously very similar to
that obtained from the standard multi-oscillator fit
technique.\cite{Prosandeev} Clearly, the three principal bands
(below 100, near 220 and 550 cm$^{-1}$) correspond to the three TO
polar modes of the average cubic structure. Each of these bands has
a tail or a bump on the high frequency side. It is natural to assign
the principal peaks to the doubly degenerate E$_1$-components and
the high frequency wings to the stiffened A$_1$-components of the
three polar modes. Similarly, the positions of the corresponding
three LO bands can be roughly read out from the plot of the
imaginary part of the inverse permittivity.

\begin{figure}[ht]
 \hspace{0cm} \centerline{\includegraphics[width=10cm, clip=true]{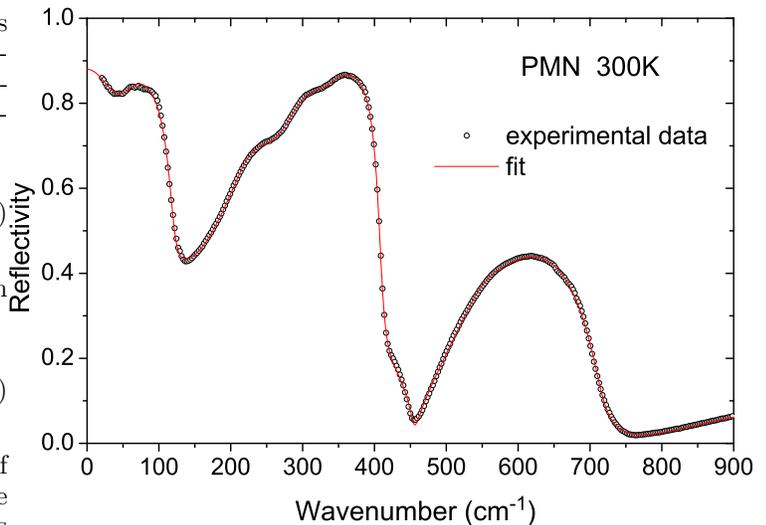}}
  \caption{(Color online) IR reflectivity spectra of the PMN single
crystal. Circles stand for the room temperature data of
Ref.~\onlinecite{Prosandeev} (by mistake denoted as 20K there), full
line stands for the fit with the model (1)-(3), parameters are given
in Table I. }
\end{figure}


A complete set of mode parameters, as obtained by adjusting the
model (eqs.~(1)-(3)) to the measured reflectivity spectrum, is given
in Table 1. The high frequency dielectric tensor was assumed
isotropic, since even in PbTiO$_3$ its two principal components
differ by less than 1~\%.\cite{Foster} It was set to
 $\epsilon_{\parallel, \infty }=\epsilon_{\perp, \infty}  =5.75$,
deduced from the available (average)
 refractive index data of PMN.\cite{McHenry}
  Furthermore, the visual agreement was much improved by
adding a weak mode near 335~cm$^{-1}$ (introduced as identical in
both $\epsilon_{\parallel}$ and $\epsilon_{\perp}$ ). The agreement
is fairly good and the number of the parameters is adequate, except
for the frequency of the overdamped E$_1$(TO1) soft mode, which
obviously cannot be reliably determined from data measured above
20~cm$^{-1}$. The value selected here makes the total phonon
contribution to the effective static dielectric permittivity of the
order of 1000, which is in qualitative agreement with our previously
published analysis.\cite{Prosandeev,KambaSM,KambaFilm}
 (There is an additional strong relaxation in the GHz frequency
region\cite{KambaSM}
 which is most likely due to the fluctuations of PNR boundaries, but it is
not in the focus of this paper.) All the parameters appear quite
realistic - the mode frequencies are reasonably close to that of
PbTiO$_3$, given for comparison in Table 1, and the trends are in
agreement with the smaller frozen polarization\cite{Grinberg} in PMN
 with respect to the spontaneous polarization of PbTiO$_3$.

\begin{table}[ht]
\centerline{
\begin{tabular}{r r r|r r}
\multicolumn{3}{c|}{PMN (this work)}&\multicolumn{2}{|c}{PbTiO$_3$
(Ref.~\onlinecite{Foster})}\\
label~~~ & $\omega$ ~~~&  $\gamma$ ~~~& label ~~~& $\omega$~~~  \\
      &         ~[cm$^{-1}$] &  [cm$^{-1}$] &  &  [cm$^{-1}$]  \\
\hline
E$_1$(TO1)& 17& 34.7&   E (TO1)  & 87.5\\
E$_1$(LO1)& 120.4& 19.8 & E (LO1)  & 128.0\\
E$_1$(TO2)& 221.4& 53.9& E (TO2)  & 218.5\\
?(LO)&334.1&56.6      &    &   \\
?(TO)&336.0&54.3         &    &   \\
E$_1$(LO2)& 409.7& 15.1& E (LO2)  & 440.5\\
E$_1$(TO3) & 545.8& 80.5& E (TO3)  & 505.0\\
E$_1$(LO3)& 714.6& 35.8& E (LO3)  & 687.0\\
\hline
A$_1$(TO1) & 63.8& 32.5&  A$_1$(TO1)  & 148.5\\
A$_1$(LO1)& 166.6& 95.1& A$_1$(LO1)& 194.0\\
A$_1$(TO2)& 284.0& 53.8 & A$_1$(TO2)& 359.5\\
?(LO)&334.1&56.6&   &           \\
?(TO)&336.0&54.3&   &               \\
A$_1$(LO2)& 461.5&4.8 & A$_1$(LO2)& 465.0\\
A$_1$(TO3)& 603.8& 117.3 & A$_1$(TO3)& 647.0\\
A$_1$(LO3)& 757.2& 160.1 & A$_1$(LO3)& 795.0\\
\end{tabular}
} \caption{Intrinsic frequencies and damping parameters
($\omega$,$\gamma$) of polar modes of PMN obtained from the fit of
the reflectivity spectrum shown in Fig.~1 with the model defined by
eqs. (1)-(3) ($\epsilon_{\parallel, \infty} =\epsilon_{\perp,
\infty} = 5.75 $), in comparison with the corresponding
 mode frequencies of room temperature  PbTiO$_3$.}
\end{table}


We do not know for sure the origin of the weak feature near
335~cm$^{-1}$. In principle, the local homogeneous polarization can
induce IR activity also for the E$_{1}$ mode derived from the
remaining F$_{\rm 2u}$ (silent) optic mode. However, the IR
dielectric strength of this mode in usual ferroelectric perovskites
is known to be extremely weak.\cite{Foster}
 Another possibility is that this spectral feature is related
 to the random occupancy on (ABO$_3$) perovskite B-sites.
As a matter of fact, there is also a weak IR active mode expected
around this frequency\cite{Prosandeev} due to the  B-site short
range ordering of NaCl-type, which activates the R$_{25}'$ Brillouin
zone corner mode of the parent cubic structure (anti-parallel
vibrations of inequivalent ions at neighboring B-sites). The
strength of this mode is determined by a compromise between its
relatively strong intrinsic strength inside of the ideally ordered
regions (Born charges at inequivalent B-sites are quite different)
and relatively small degree of this B-site order. This mode was in
fact quite clearly seen in IR spectra of compounds with a larger
degree of B-site order, like PST and similar systems with 1:1 B-site
stoichiometry, \cite{Setter-Petzelt,Buixaderas} so that it is
perhaps the most likely interpretation.

\begin{figure}[ht]
 \hspace{0cm} \centerline{\includegraphics[width=
10cm, clip=true]
    {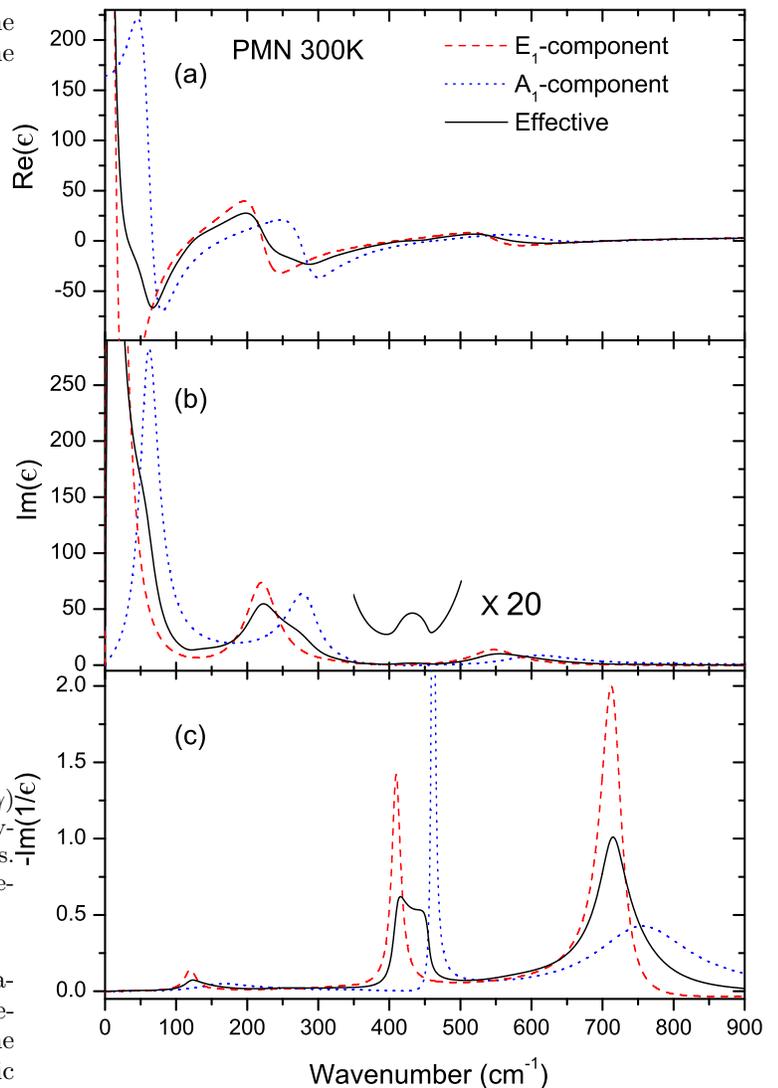}}
     \caption{(Color online) Spectra of real (a) and imaginary (b) part
of the permittivity and of imaginary part of the inverse
permittivity (c) calculated from the model defined by eqs. (1)-(3)
with the adjusted parameters given in Table I. Full line corresponds
to the effective macroscopic permittivity $\epsilon_{\rm eff}$,
dashed and dotted corresponds to the $\epsilon_{\perp}$ and
$\epsilon_{\parallel}$ components of the local permittivity,
respectively. }
\end{figure}


The imaginary part of effective permittivity (Fig.~2 (b)) seems to
indicate another weak but clear mode with TO frequency near 400
cm$^{-1}$. The origin of this mode was a puzzle since realistic
calculations\cite{Prosandeev} show that no TO mode is expected
around this frequency. Interestingly, the effective medium model
proposed here reproduces this band without assuming any intrinsic TO
frequency near 400 cm$^{-1}$. Such modes, so called geometrical
resonances, are actually interfacial modes intimately related to the
heterogeneity of the medium, and they are known for example from IR
spectra of ceramics of anisotropic materials.\cite{GRPecharroman}
This purely geometric resonance related to the double reflectivity
minimum near 400~cm$^{-1}$ appears to be pronounced due to a
relatively large splitting of the LO2 mode in comparison with its
damping, and it is a generic feature present in many other relaxor
perovkites.

A systematic analyses of IR reflectivity data of other relaxor
perovskites as well as of temperature dependences within this model
are now in progress. In PMN, the frequency of the A$_1$(TO1) mode
increases on temperature lowering and it is obvious that this mode
coincides with that which was denoted soft mode in the previous low
temperature IR\cite{KambaSM,Prosandeev,KambaFilm} and neutron
studies.\cite{INSSM} On the contrary, the E$_1$-component of TO1
mode seems to remain rather soft in all perovskite relaxors.
Actually, it would be extremely helpful to develop an appropriate
averaging scheme also for the polar modes in Raman scattering and
inelastic neutron scattering spectra, in order to form a solid basis
for comparison. For example, it seems to us that Raman bands
denoted\cite{Toulouse} as A-F could be related to the polar modes
A$_1$(TO1), A$_1$(LO3), LO1, TO2 + silent, TO3 and LO2 modes,
respectively. At the present stage however, a direct quantitative
comparison is difficult and could be even misleading.

The present model can be modified in many ways. For example, one may
consider an elongation of PNR along the direction of the frozen
polarization or to consider the contribution of the residual
non-polar matrix, both possible using the analogical effective
medium schemes developed for ceramics\cite{PRBPecharroman}. More
interestingly, one can hope to incorporate a more realistic
information about the geometry of PNR structure within a more
sophisticated effective medium approach. At the level of the crude
approximation (eq. 1), it actually does not matter whether the
directions of the frozen polarization are completely random or
whether they tend to be oriented preferentially along a family of
high symmetry directions (most likely, 111 or 100). None of these
modifications is expected to have a drastic influence on the
intrinsic phonon frequencies as determined here, but it is possible
that it may explain the spectra assuming  smaller damping
parameters.

In conclusion, IR spectroscopy brings evidences of splitting of
polar optic modes in perovskite relaxors. We deduce that at phonon
frequencies and at the length-scale of PNR's, the dielectric
function is strongly anisotropic. It is demonstrated on the PMN
crystal case that a simple effective medium model can reproduce the
average permittivity and reflectivity spectra fairly well. The
effective medium approach also explains the pseudo- TO mode near
400~cm$^{-1}$ as a consequence of nanoscopic heterogeneity.
 It is believed that the model can be used for systematic estimating
  the frequencies  of the three principal polar modes in perovskite
 relaxors
and  their A$_1$-E$_1$ splitting due to the underlying polarization,
 provided  the reflectivity data
are known in a sufficiently broad interval (20-1000~cm$^{-1}$).

\begin{acknowledgments} We would like to thank   C.
Pecharrom\'{a}n from Instituto de Ciencia de Materiales, Consejo
Superior de Investigaciones Cientificas, Madrid  and  I.
Rychetsk\'{y} from Institute of Physics  ASCR, Praha for stimulating
discussions. The work has been supported by the Czech Academy of
Sciences (projects  A1010213 and AVOZ1-010-914) and by the Grant
Agency of the Czech Republic (project 202/04/0993).
\end{acknowledgments}

\end{document}